\documentclass[aps,preprint,prd,showpacs,nofootinbib]{revtex4}

\usepackage{latexsym}
\usepackage{CJK}
\usepackage{amsmath}
\usepackage{graphicx}
\usepackage{dcolumn}
\usepackage{bm}
\usepackage{amssymb}
\usepackage{latexsym}
\usepackage{color}
\usepackage[colorlinks,linkcolor=magenta,anchorcolor=blue,citecolor=blue]{hyperref}

\def\be{\begin{equation}}
\def\ee{\end{equation}}
\def\ba{\begin{eqnarray}}
\def\ea{\end{eqnarray}}

\begin{document}

\title{Pre-inflationary primordial perturbations }

\author{Yong Cai$^{1,}$\footnote{caiyong13@mails.ucas.ac.cn}}
\author{Yu-Tong Wang$^{1}$\footnote{wangyutong12@mails.ucas.ac.cn}}
\author{Yun-Song Piao$^{1,2}$\footnote{yspiao@ucas.ac.cn}}

\affiliation{$^1$ School of Physics, University of Chinese Academy of
Sciences, Beijing 100049, China}

\affiliation{$^2$ State Key Laboratory of Theoretical Physics, Institute of Theoretical Physics, \\
Chinese Academy of Sciences, P.O. Box 2735, Beijing 100190, China}

\begin{abstract}

The large-scale power deficit in the cosmic microwave background fluctuations might be
relevant with the physics of pre-inflation, a bounce, or a
superinflationary phase preceding slow-roll inflation, which can
provide a singular-free realization of inflation. We investigate
the primordial perturbations from such pre-inflationary
evolutions, which generally may consist of multiple phases with
different background dynamics, and give a universal formula for
the power spectrum of primordial perturbations in terms of the
recursive Bogoliubov coefficients. We also apply our formula to
corresponding cases  and show how the intensity of large-scale
power suppression is affected by the pre-inflationary physics.

\end{abstract}
\maketitle

\section{Introduction}

As the paradigm of the early universe, inflation has been
generally regarded as a possible solution of the horizon,
flatness, entropy, homogeneity, isotropy and primordial monopole
problems\cite{Guth:1980zm},\cite{Starobinsky:1980te},\cite{Linde:1981mu},\cite{Albrecht:1982wi}.
But maybe far more attractive is that inflation can generate the
primordial perturbations, which have grown into all the structures
observed in our universe today. The observations of cosmic microwave background (CMB) by Planck
and WMAP has provided us with more and more information of the
early universe, which shows that the single field slow-roll
inflationary model is more likely to be the right one.

However, a large-scale (or low-$l$) power deficit in CMB TT-mode
spectrum observed by WMAP \cite{Spergel:2003cb} and recently
confirmed by the Planck Collaboration\cite{Ade:2013xsa}\cite{Ade:2013uln} with
higher precision is not concordant with the standard slow-roll
inflation. It is hard to attribute this power deficit to the
foreground as it has been observed by experiments with higher and
higher statistical significance. Though the cosmic variance could
be a source of this deficit, it is still very likely that the
large-scale anomalies are induced by the physics preceding
inflation, as the larger are the scales of the perturbations, the
earlier are the times corresponding to their horizons exiting.

The slow-roll inflation might last for just the minimal number of
$e$-folds, i.e., just enough \cite{Ramirez}, and thus the power
deficit on a large scale may be attributed to the pre-inflationary
non-slow-roll evolution. In this case, the Planck best-fit
single-field inflationary model only actually provides a fit for
the intermediate and small angular scales. After the WMAP1 data
were released, some studies have been done in
Refs.\cite{Contaldi:2003zv},\cite{Cline:2003ve},\cite{Piao:2003zm},\cite{Piao:2003hh},\cite{Powell:2006yg},\cite{Falciano:2008gt},\cite{Mielczarek:2008pf}
along this line, and also recently
\cite{Liu:2013kea},\cite{Biswas:2013dry},\cite{Liu:2013iha},\cite{Dudas:2012vv},\cite{Kitazawa:2014dya},\cite{Qiu:2014nla},\cite{Liu:2010fm},\cite{BouhmadiLopez:2012by};
see Ref.\cite{Cicoli:2014bja} for a review.

However, it might be more interesting that the large-scale
anomalies could be relevant with the physics solving the initial
singularity problem of inflation
\cite{Piao:2003zm},\cite{Dudas:2012vv}. In the bouncing model (see,
e.g., \cite{Battefeld:2014uga},\cite{Lehners:2011kr} for reviews),
initially the universe is in a contracting phase, and then it
bounces into an expanding phase, which results in a solution to
the cosmological singularity problem. In
Refs.\cite{Piao:2003zm},\cite{Falciano:2008gt},\cite{Mielczarek:2008pf},\cite{Liu:2013kea},\cite{Biswas:2013dry},\cite{Qiu:2014nla},
it was noticed that if the universe is initially in a contracting
phase and after the bounce it begins to inflate, the power
spectrum of primordial perturbations will get a large-scale
cutoff, which may naturally explain the power deficit of the CMB
TT-mode spectrum at low-$l$; see, e.g., \cite{Liu:2013kea} for
details. In addition, it was also observed in \cite{Wang:2014abh}
that if the pre-inflationary bounce actually occurs, the BB-mode
correlation at low-$l$ is also suppressed, while the TB- and
EB-mode correlations on a corresponding scale may be enhanced.

The superinflationary phase before slow-roll inflation may also
provide a singular-free realization of inflation, which in the meantime
explains the anomalies of the CMB power spectrum
\cite{Liu:2013iha},\cite{Labrana:2013oca}; see also
\cite{Liu:2014tda},\cite{Pirtskhalava:2014esa} for an almost flat
pre-inflationary universe.

In the above pre-inflationary scenarios,
e.g., \cite{Piao:2003zm},\cite{Liu:2013kea},\cite{Liu:2013iha},
initially the primordial perturbation is deep inside the horizon,
which naturally set itself in Bunch-Davies(BD) vacuum. This
implies that we can calculate a large-scale power spectrum without
any assumption for the initial state of primordial perturbations.
However, it is possible that the pre-inflationary era might
consist of multiple phases with different background evolution,
e.g., \cite{Cicoli:2014bja}. In a certain sense, the introduction of
multiple pre-inflationary phases may better simulate the physics
of the pre-inflationary era, since due to the complexity of
pre-inflationary physics, sometimes a single phase can hardly
reflect the drastic change of the background parameters,
e.g., \cite{Kitazawa:2014dya}.
Therefore, it is interesting to have a quantitative estimate for
the power spectra of primordial perturbations from an arbitrary
pre-inflationary era, involving the multiple phases with different
background dynamics.

This paper is organized as follows. In Sec.\ref{au}A, we introduce
the evolution of the pre-inflationary background, which we will focus
on, consisting of multiple phases. We require that the primordial
perturbations can be produced in these phases. In Sec.\ref{au}B,
we perform a model-independent calculation for the primordial
perturbations and give a universal formula for the power spectra
of primordial perturbations in terms of the recursive Bogoliubov
coefficients. In Sec.\ref{double}, we apply our formula to the
bounce inflation and the superinflation preceding slow-roll
inflation  and show how the intensity of the large-scale power
suppression of a primordial spectrum is affected by the
pre-inflationary physics. We will see that a large-scale suppressed
primordial spectrum may result in the power deficit at low-$l$ in
the CMB TT-mode spectrum; however, the intensity of the power
suppression is model-dependent. Sec.\ref{conclusion} is the
discussion.

In Appendix A, we will investigate the Wronskian constraint for
primordial perturbations  and argue that in a certain sense the
scenario with pre-inflationary era is equivalent to the inflation
scenario with the non-Bunch-Davis initial state. In Appendix B, we
will approximately estimate the spectra index of primordial
perturbations produced in each phase.

\section{Pre-inflationary primordial perturbations}\label{au}

\subsection{The pre-inflationary background}

The pre-inflationary era may consist of multiple phases. We define
the inflation as   phase 0, the latest pre-inflationary
phase as phase $1$, and so on.

The cosmological evolution of phase $i$ is \be a_i\sim
\eta^{\frac{2}{1+3\omega_i}},\,\,\,{or}\,\,\,(-\eta)^{\frac{2}{1+3\omega_i}},\label{aa}\ee
where $\omega_i=p_i/\rho_i$ is a constant, $\omega_i\neq
-{1}/{3}$, and $\eta=\int dt/a$ is the conformal time. The phase
$i$ may be expanding or contracting.

However, noting the continuities of $a$ and $|H|$, the evolutions
of the inflation and  phase $i$ can be rewritten as
\be a_0(\eta)=\frac{a_0(\eta_0)}{1-{\cal
H}_{inf}\cdot(\eta-\eta_0)} ,\ee \be
a_i(\eta)=a_{i}(\eta_{i-1})\left[1+ \frac{1+3\omega_i}{2}{\cal
H}_i(\eta_{i-1})\cdot
(\eta-\eta_{i-1})\right]^{\frac{2}{1+3\omega_i}} \label{a},\ee
respectively, where $\eta_0$ is that at the onset of inflation,
${\cal H}_{inf}={\cal H}_0(\eta_0)$ is the conformal Hubble
parameter during slow-roll inflation, and $\eta_{i-1}$ is the
matching time between phase $i$ and phase $i-1$, which signals the
onset of phase $i-1$.

It seems that $\omega_i$ is arbitrary. However, if we require that
initially all perturbation modes are in the BD state, which is right
only if its wavelength $\lambda\ll {1/{\cal H}}$, $\omega_i$ will
be constrained. That the perturbation mode with $\lambda\sim 1/k$
extends outside the horizon in phase $i$ marks the primordial
perturbation that is produced in this phase, which requires ${\cal
H}_i(\eta_i)<k<{\cal H}_i(\eta_{i-1})$. Thus $|{\cal H}_i(\eta)|$
must increase with
time, which implies \ba \omega_i & < & -1/3~\,\,\,\, for\,\, {{\rm the}}\,\, {{\rm expanding}}\,\, {{\rm phase}},\nonumber \\
or \,\,\,\omega_i & > & -1/3~ \,\,\,\, for\,\, {{\rm the}} \,\,
{{\rm contracting}}\,\, {{\rm phase}}. \label{w}\ea Or equally in
other words, (\ref{w}) must be satisfied to guarantee the
primordial perturbations that can be produced during phase $i$. The
slow-roll inflation is the evolution with $\omega\simeq -1$, which
satisfies (\ref{w}).

In general, there may be multiple pre-inflationary phases. Any one
of them might be a contracting phase or an expanding phase.
Therefore, there are two kinds of typical scenarios which have
aroused lots of interests, i.e., the bouncing scenario and the
superinflation scenario,  which we will focus on in Secs. III.A
and III.B.

In the bouncing scenario, a bounce happened near the end of the
contracting phase  then was followed by an expanding phase with
$\omega<-1/3$. The bounce may be implemented by applying a
higher-order derivative field
\cite{Qiu:2011cy},\cite{Easson:2011zy},\cite{Osipov:2013ssa},\cite{Koehn:2013upa},
which is ghost-free, and also viscous fluid
\cite{Myrzakulov:2014hva} and modified gravity
\cite{Bamba:2013fha},\cite{Biswas:2005qr},\cite{Calcagni:2010bj},\cite{Paul:2014cxa},\cite{Oikonomou:2014yua}.

The expansion with $\omega<-1$ is called superinflation,
which is similar to the emergent scenario and describes a
monotonically expanding universe with increasing energy density.
The primordial perturbations generated during the superinflation
have been studied earlier in
Refs.\cite{Piao:2004tq},\cite{Baldi:2005gk}. The case with
$\omega\ll -1$ corresponds to the slow expansion scenario, which
has been proposed in Ref.\cite{Piao:2003ty} (see also
\cite{CNT},\cite{Hinterbichler:2012fr} for the genesis scenario) and
investigated in detail in
Refs.\cite{Piao:2010bi},\cite{Liu:2011ns}. It is shown in
Ref.\cite{Liu:2011ns} that there is no ghost instability during
superinflation. Thus the pre-inflationary universe may also be
superinflating. It is interesting to notice that both bounce and
superinflation preceding slow-roll inflation may provide a
singular-free realization of inflation.

However, if the pre-inflationary era is the expanding phase with
$\omega=1$ characterized by fast-rolling dominance, e.g.,
\cite{Contaldi:2003zv},\cite{Cicoli:2013oba},\cite{Pedro:2013pba},\cite{Jain:2008dw},
except for the initial fast-roll inflation, e.g., \cite{Hazra:2014jka},
or the expanding phase with radiation dominance
\cite{Cline:2003ve},\cite{Powell:2006yg}, initially the
perturbation mode should be outside the horizon, so it is not
clear how to set its initial condition. It is possible that its
initial value is set in a phase preceding fast-rolling dominance;
however, this phase is still required to satisfy (\ref{w}), e.g., a
higher energy inflation preceding the fast-rolling phase
\cite{Namjoo:2012xs}; see \cite{Cicoli:2014bja} for comments, and see
also \cite{BouhmadiLopez:2012by} for a domain wall or cosmic
string phase. Here, we will not involve this issue  and will
assume that all pre-inflationary phases satisfy (\ref{w}), which
assures initially all perturbation modes are naturally set in BD
vacuum and each phase may contribute to the
production of primordial perturbations with the
certain range of wave numbers.

\begin{figure}[t!]
    \centering
\begin{minipage}[b]{0.6\linewidth}
    \centering
    \includegraphics[width=10.0cm]{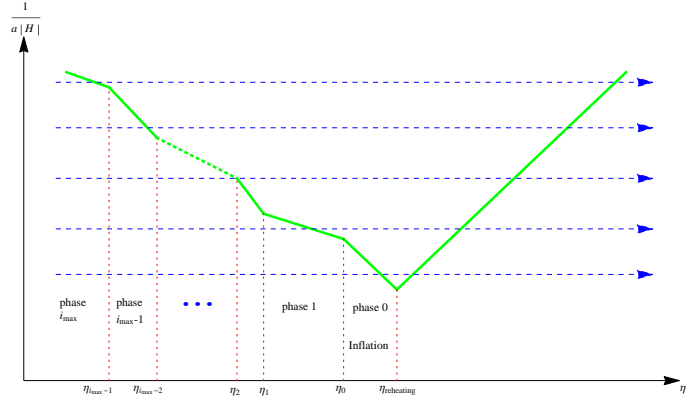}
    \end{minipage}
    \hspace{0.05cm}
\caption{The green line denotes the evolution of the comoving
Hubble radius $\frac{1}{a|H|}$, and $\eta_i$ signals the onset of
phase $i$. The blue dashed lines denote the evolutions of the
primordial perturbations.} \label{fig:evo}
\end{figure}

We qualitatively plot the evolutions of background and
perturbation modes, which we will focus on, in Fig.\ref{fig:evo}.

\subsection{The power spectrum of pre-inflationary perturbations}

The equation of the primordial perturbation $\cal R$ is
\cite{Muk},\cite{KS} \be u^{\prime\prime}+\left
(c_s^2k^2-\frac{z^{\prime\prime}}{z}\right)u=0 \label{u},\ee where
$u\equiv z{\cal R}$, the prime is the derivative with respect to
conformal time, $z\equiv a\sqrt{2M_p^2\epsilon}/c_s$ \cite{GM},
and $z\simeq a\sqrt{2M_p^2 |\epsilon|}/c_s$ \cite{Liu:2011ns} for
the superinflationary phase, and the definition of $\epsilon$ is
$\epsilon\equiv -{\dot H}/{H^2}={3(1+\omega)\over 2}$. We assume
$c_s^2=1$ for simplicity.

In the slow-roll inflationary phase, we have \be
\frac{z^{\prime\prime}_0}{z_0}\simeq \frac{2{\cal H}_{inf}^2
}{\left[1-{\cal H}_{inf}(\eta-\eta_0) \right]^2}, \ee the solution
of Eq.(\ref{u}) is \ba u_{0} & = & \sqrt{-k{\eta}_{eff_0}} \left[
C_{0,1}H^{(1)}_{3/2}\left(-k{\eta}_{eff_0}\right)
+C_{0,2}H^{(2)}_{3/2}\left(-k{\eta}_{eff_0}\right)
\right],\label{u0}\ea where $H^{(1)}_{3/2}$ and $H^{(2)}_{3/2}$
are the $(3/2)$th order Hankel functions of the first and second
kinds, respectively, and ${\eta}_{eff_0}=\eta-\eta_0-\frac{1}{{\cal
H}_{inf}}$.

In pre-inflationary phase $i$, we have \be
\frac{z_i^{\prime\prime}}{z_i} \simeq\frac{(1-3\omega_i){\cal
H}_i^2(\eta_{i-1})}{2\left[1+\frac{1+3\omega_i}{2}{\cal
H}_i(\eta_{i-1})\cdot(\eta-\eta_{i-1})\right]^2},  \ee the
solution of Eq.(\ref{u}) is \ba & & u_{i}  = \sqrt{-k{\eta}_{eff_i}}
\left[C_{i,1} H^{(1)}_{v_i}\left(-k{\eta}_{eff_i}\right) + C_{i,2}
H^{(2)}_{v_i}\left(-k{\eta}_{eff_i}\right) \right] , \label{ucc}\ea
where $
v_i=\frac{3}{2}\left|\frac{1-\omega_i}{1+3\omega_i}\right|$.
$H^{(1)}_{v_i}$ and $H^{(2)}_{v_i}$ are the $v_i$th order Hankel
functions of the first and second kinds, respectively, and
${\eta}_{eff_i}=\eta-\eta_{i-1}+\frac{2}{(1+3\omega_i){\cal
H}_i(\eta_{i-1})}$.
In fact, Eq.(\ref{u0}) can be obtained from Eq.(\ref{ucc}) while $i=0$, but $\eta_{i-1}$ should be replaced by $\eta_{0}$.

Here, we require that around and at the matching surface besides
there is not the ghost instability; there is also not gradient
instability, i.e., $c_s^2>0$. In this case, the perturbation can
continuously pass through the matching surface between two
adjacent phases, and its spectrum is insensitive to the physical
details around the matching surface; see
e.g., \cite{Battarra:2014tga} for the bounce. The coefficients $C_i$
in Eqs.(\ref{u0}) and (\ref{ucc}) are determined by requiring
the continuity of $u$ and $u^{\prime}$ at the matching surface. We
can write the coefficients $C_{i,1}$ and $C_{i,2}$ of phase $i$
recursively as \ba \left(
\begin{array}{ccc} C_{i,1}\\C_{i,2}
\end{array}\right)&=&{\cal
M}^{(i,i+1)}\times\left(\begin{array}{ccc}C_{i+1, 1}\\C_{i+1,
2}\end{array}\right) \label{rec}\nonumber\\&=&{\cal
M}^{(i,i+1)}\times{\cal M}^{(i+1,i+2)}\times\cdots\times{\cal
M}^{(i_{max}-1,i_{max})}\times \left( \begin{array}{ccc}
C_{i_{max},1}\\C_{i_{max},2}\end{array}\right) ,\label{recur}\ea
where the earliest phase is defined as the phase $i_{max}$, and
${\cal M}^{(i,i+1)}=\left(\begin{array}{ccc} {\cal M}_{11}&{\cal
M}_{12}\\{\cal M}_{21}&{\cal M}_{22}\end{array}\right)$ is the
recursive matrix, which is given by \ba {\cal M}_{11}&=&\frac{i\pi
\sqrt{x y}
}{8}\bigg\{\left[-H^{(1)}_{-1+v_{i+1}}(y)+H^{(1)}_{1+v_{i+1}}(y)\right]
H^{(2)}_{v_i}(x)+\left[H^{(2)}_{-1+v_i}(x)-H^{(2)}_{1+v_i}(x)\right]H^{(1)}_{v_{i+1}}(y)
\label{ditui} \nonumber\\
&+&(x^{-1}-y^{-1})
H^{(2)}_{v_i}(x)H^{(1)}_{v_{i+1}}(y)\bigg\},
\nonumber\\
{\cal M}_{12}&=&\frac{i\pi \sqrt{x y}
}{8}\bigg\{\left[-H^{(2)}_{-1+v_{i+1}}(y)+H^{(2)}_{1+v_{i+1}}(y)\right]
H^{(2)}_{v_i}(x)+\left[H^{(2)}_{-1+v_i}(x)-H^{(2)}_{1+v_i}(x)\right]H^{(2)}_{v_{i+1}}(y)
\nonumber\\
&+&(x^{-1}-y^{-1})
H^{(2)}_{v_{i}}(x)H^{(2)}_{v_{i+1}}(y)\bigg\},
\nonumber\\
{\cal M}_{21}&=&\frac{i\pi \sqrt{x y}
}{8}\bigg\{\left[H^{(1)}_{-1+v_{i+1}}(y)-H^{(1)}_{1+v_{i+1}}(y)\right]
H^{(1)}_{v_i}(x)-\left[H^{(1)}_{-1+v_i}(x)-H^{(1)}_{1+v_i}(x)\right]H^{(1)}_{v_{i+1}}(y)
\nonumber\\
&-&(x^{-1}-y^{-1})
H^{(1)}_{v_{i}}(x)H^{(1)}_{v_{i+1}}(y)\bigg\},
\nonumber\\
{\cal M}_{22}&=&\frac{i\pi \sqrt{x y}
}{8}\bigg\{\left[H^{(2)}_{-1+v_{i+1}}(y)-H^{(2)}_{1+v_{i+1}}(y)\right]
H^{(1)}_{v_i}(x)-\left[H^{(1)}_{-1+v_i}(x)-H^{(1)}_{1+v_i}(x)\right]H^{(2)}_{v_{i+1}}(y)
\nonumber\\
&-&(x^{-1}-y^{-1}) H^{(1)}_{v_{i}}(x)H^{(2)}_{v_{i+1}}(y)\bigg\},
\ea where $x =-\frac{2k}{(1+3\omega_i){\cal H}_i(\eta_i)}$ and $ y
=-\frac{2k}{(1+3\omega_{i+1}){\cal H}_{i+1}(\eta_i)}$.
When $i=0$, $\eta_{i-1}$ should be replaced with $\eta_0$. A result similar to (\ref{ditui}) was obtained in \cite{Cicoli:2014bja}.

In the earliest phase, the coefficients $C_{i_{max},1}$ and
$C_{i_{max},2}$ are determined by the initial condition. When
$k^2\gg\frac{z_{i_{max}}^{\prime\prime}}{z_{i_{max}}}$, the
perturbation mode is deep inside the horizon, which is set in BD
vacuum, \be u\sim\frac{1}{\sqrt{2k}}e^{-ik\eta}.\label{uggk}\ee
When $k^2\gg\frac{z_{i_{max}}^{\prime\prime}}{z_{i_{max}}}$,
$u_{i_{max}}$ given in Eq.(\ref{ucc}) should approximate to the
form in Eq.(\ref{uggk}).
Thus, we get \be C_{i_{max},1}=\frac{\sqrt{\pi}}{2\sqrt{k}},\,\,\,
C_{i_{max},2}=0. \ee Obviously, $C_{i_{max},1}$ and
$C_{i_{max},2}$ satisfy the so-called Wronskian (or canonical
normalization) constraint (see e.g. \cite{Ashoorioon:2006wc}, \cite{Ashoorioon:2010xg}) \be  \frac{4k}{\pi}\left(
|C_{i_{max},1}|^2-|C_{i_{max},2}|^2  \right)=1. \ee And actually,
for the following phase $i$, $C_{i,1}$ and $C_{i,2}$ will always
satisfy the Wronskian constraint, which will be proved in Appendix
A. $C_{i,1}$ and $C_{i,2}$ are related to the so-called Bogoliubov coefficients by Eq.(\ref{bogoliubov}).


The power spectrum of $\cal R$ is \be {\cal P}_{\cal
R}=\frac{k^3}{2\pi^2}\left|\frac{u_{0}}{z_0}\right|^2 .\label{pu}
\ee After substituting Eq.(\ref{u0}) into Eq.(\ref{pu}) and
requiring $\left|k(\eta-\eta_0)-\frac{k}{{\cal H}_{inf}}\right|\ll 1$, we
get a universal formula \be {\cal P}_{\cal R}={\cal P}^{inf}_{\cal
R}\frac{4}{\pi}k\left|C_{0,1}-C_{0,2}\right|^2, \label{prc1c2}\ee
where ${\cal P}^{inf}_{\cal
R}=\frac{1}{2M_p^2\epsilon_{inf}}\left(\frac{H_{inf}}{2\pi}\right)^2$
is the spectrum of the standard slow-roll inflation, $H_{inf}={
{\cal H}_0(\eta_0)}/{a_0(\eta_0)}$ is the Hubble parameter during
inflation, and in this sense actually ${\cal H}_0(\eta_0)={\cal
H}_{inf}$. The spectral index of $\cal R$ is \be n_{\cal
R}=n_{inf}+\frac{d\ln(k\left|C_{0,1}-C_{0,2}\right|^2)}{d\ln
k},\label{nr}\ee where $n_{inf}={d\ln{\cal P}^{inf}_{\cal R}\over
d\ln{k}}+1$ is the spectral index of slow-roll inflation, which is
nearly unity.

The coefficients $C_{0,1}$ and $C_{0,2}$ are determined by the
recursive Eq.(\ref{recur}), and thus the effects of all
pre-inflationary phases are encoded in $C_{0,1}$ and $C_{0,2}$,
which are nontrivial. By calculating Eq.(\ref{prc1c2}), it can be
found that the perturbations produced in phase $i$ roughly have a
power spectrum \be {\cal P}_{\cal R}\sim k^{3-2v_i}.
\label{pnr}\ee We give a proof for this result in Appendix B. It
is observed in Appendix B that the power spectrum of perturbations
is Eq.(\ref{pnr}) but modulated with a small oscillation, which is
induced by the evolution of perturbation through the matching
surface between adjacent phases.

\section{The large-scale power suppression }\label{double}

In this section, we will apply our universal formula
(\ref{prc1c2}) to the bounce inflation and the superinflation
preceding slow-roll inflation, and show how the intensity of the
large-scale power suppression in the CMB fluctuations is affected
by the pre-inflationary evolution.

The large-scale power suppression requires that the $e$-folding
number of slow-roll inflation is just enough or less, and in the
meantime the pre-inflationary era can contribute a strong
blue-tilt spectrum. The spectral index of pre-inflationary
perturbations is approximately \be n_{\cal R}-1\simeq
\frac{d\ln(k\left|C_{0,1}-C_{0,2}\right|^2)}{d\ln k}.
\label{nr1}\ee Thus the intensity of suppression can be
model dependent, since $|C_{0,1}-C_{0,2}|^2$ is different for
different parameters $\omega_i$ and $\eta_i$.

Here, we will focus on the cases that the pre-inflationary era
consists of two phases, i.e., $i_{max}=2$, and set ${\cal
H}_0(\eta_0)={\cal H}_{inf}$ and $\eta_0=0$ is the time at the
onset of inflation; thus, we have $\eta<0$ in the pre-inflationary
phases and $\eta>0$ in the inflationary phase. According to
Sec.\ref{au}B, for $i_{max}=2$, the solutions of Eq.(\ref{u}) are
\ba
u_0&=&\sqrt{-k\eta+\frac{k}{ {\cal H}_{inf}}}\left[C_{0,1}H^{(1)}_{3/2}(-k\eta+\frac{k}{ {\cal H}_{inf}})+C_{0,2}H^{(2)}_{3/2}(-k\eta+\frac{k}{ {\cal H}_{inf}  } )\right],\label{u012}\nonumber\\
u_1&=&\sqrt{-k\eta-\frac{2k}{(1+3\omega_1){\cal H}_1(0)}}\Big[ C_{1,1}H^{(1)}_{v_1}(-k\eta-\frac{2k}{(1+3\omega_1){\cal H}_1(0)})\nonumber\\
&\,&\qquad\qquad\qquad\qquad\qquad\quad\,\, +C_{1,2}H^{(2)}_{v_1}(-k\eta-\frac{2k}{(1+3\omega_1){\cal H}_1(0)})  \Big],\nonumber\\
u_2&=&\frac{\sqrt{\pi}}{2}\sqrt{-\eta+\eta_1-\frac{2}{(1+3\omega_2){\cal
H}_2(\eta_1)}}H^{(1)}_{v_2}(-k\eta+k\eta_1-\frac{2k}{(1+3\omega_2){\cal
H}_2(\eta_1)}), \label{set}\ea  respectively, where $\eta_1<0$ is
the conformal time at the matching surface between phase $1$ and
phase $2$.

We define \be P(\eta_1,\omega_1,\omega_2)=\frac{ {\cal P}_{\cal R}
}{ {\cal P}_{\cal R}^{inf} } =\frac{4}{\pi}k|C_{0,1}-C_{0,2}|^2,
\label{pnew}\ee where
according to Eq.(\ref{rec}), we have \ba \left(
\begin{array}{ccc} C_{0,1}\\C_{0,2} \end{array}\right)={\cal
M}^{(0,1)}\times{\cal M}^{(1,2)}\times \left( \begin{array}{ccc}
\frac{\sqrt{\pi}}{2\sqrt{k}}\\ 0\end{array}\right) ,\ea and the
components of ${\cal M}^{(0,1)}$, ${\cal M}^{(1,2)}$ can be
obtained from Eq.(\ref{ditui}). $P(\eta_1,\omega_1,\omega_2)$
reflects the shape of the spectrum. We will check how the shape of the spectrum, as well as the intensity of suppression on a large scale,
changes with the parameters $\eta_1$, $\omega_1$, and $\omega_2$ in
the bounce inflation and the superinflation preceding slow-roll
inflation.

\subsection{The superinflationary phase before inflation}\label{eei}

The superinflation is defined as the evolution with
$\epsilon\equiv -{\dot H}/H^2<0$ or $\omega<-1$,
e.g., \cite{Piao:2004tq},\cite{Baldi:2005gk},\cite{Capozziello:2005tf},\cite{Piao:2003ty}.
The model with one single superinflationary phase
$\omega=-{5}/{3}$ preceding slow-roll inflation has been built
in string theory in Ref.\cite{Liu:2013iha}. In
Refs.\cite{Liu:2014tda},\cite{Pirtskhalava:2014esa}, the
pre-inflationary universe is in a slowly expanding genesis phase
with $\epsilon\ll -1$, which is almost Minkowski space. This
genesis phase actually also belongs to the superinflation, but
since $\epsilon\ll -1$, the expansion is actually slow; see also
\cite{Labrana:2013oca} for the emergent universe.

We will fix phase 2 with $\omega_2=-{5}/{3}$. Thus in a certain
sense phase 1 actually corresponds to an intermediate phase,
which obviously must also be expanding. 
The duration that this intermediate phase lasts is $\Delta
\eta=|\eta_1|$. Here, the parameter $\omega_1$ in phase 1 is
model-dependent but should satisfy $\omega_1<-{1}/{3}$. By
requiring the continuity of $|{\cal H}|$, we get \be {\cal
H}_1(0)={\cal H}_{inf} \,,\qquad {\cal H}_2(\eta_1)=\frac{2{ \cal
H}_{inf} }{2+(1+3\omega_1)\eta_1{\cal H}_{inf}}, \label{h12}\ee
where ${\cal H}_{inf}$ is determined by the amplitude of CMB
fluctuations. Thus in equation set (\ref{set}) only parameters
$\eta_1$ and $\omega_1$ are left to be free.

\begin{figure}[t!]
    \centering
\begin{minipage}[b]{0.45\linewidth}
    \centering
    \includegraphics[width=0.95\textwidth]{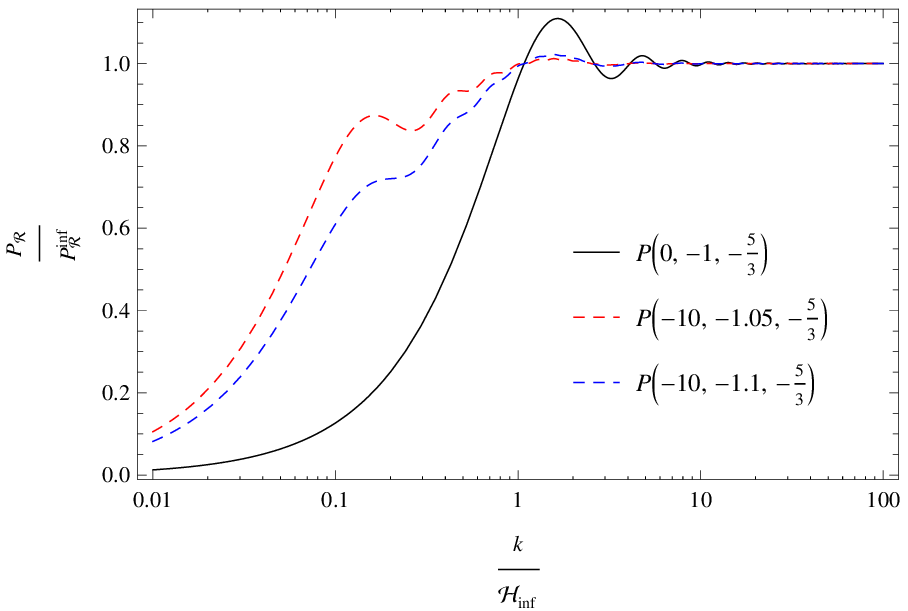}
    \end{minipage}
    \hspace{0.05cm}
\begin{minipage}[b]{0.45\linewidth}
    \centering
    \includegraphics[width=0.95\textwidth]{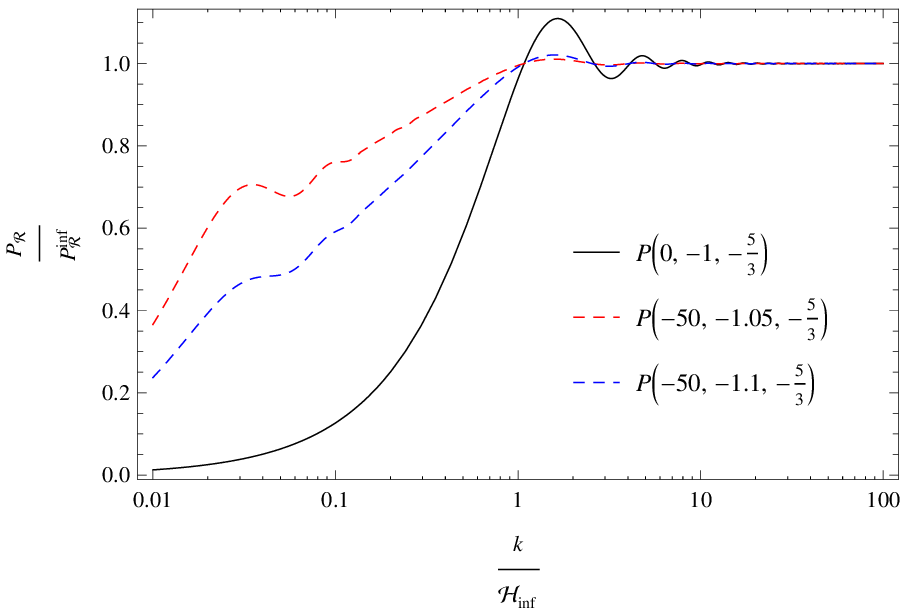}
    \end{minipage}
    \hspace{0.05cm}
\caption{The power spectrum $P(\eta_1,\omega_1,-{5}/{3})$ of the
model, in which there is an intermediate phase between
superinflation and slow-roll inflation, with different $\eta_1$
and $\omega_1$. $P(0,-1,-{5}/{3})$ corresponds to the case in
Ref.\cite{Liu:2013iha} without the intermediate phase.}
    \label{fig:eei}
\end{figure}

\begin{figure}[t!]
    \centering
\begin{minipage}[b]{0.45\linewidth}
    \centering
    \includegraphics[width=0.95\textwidth]{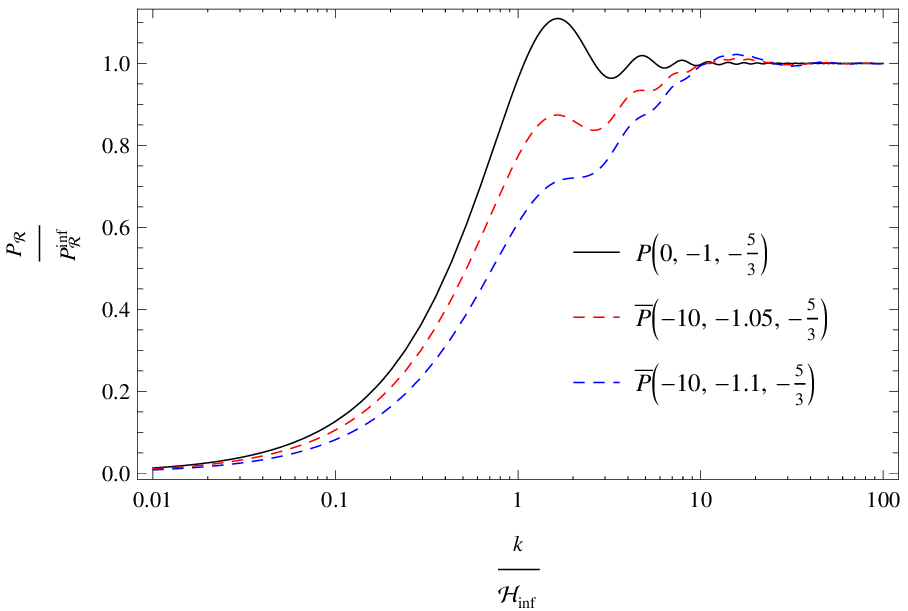}
    \end{minipage}
    \hspace{0.05cm}
\begin{minipage}[b]{0.45\linewidth}
    \centering
    \includegraphics[width=0.95\textwidth]{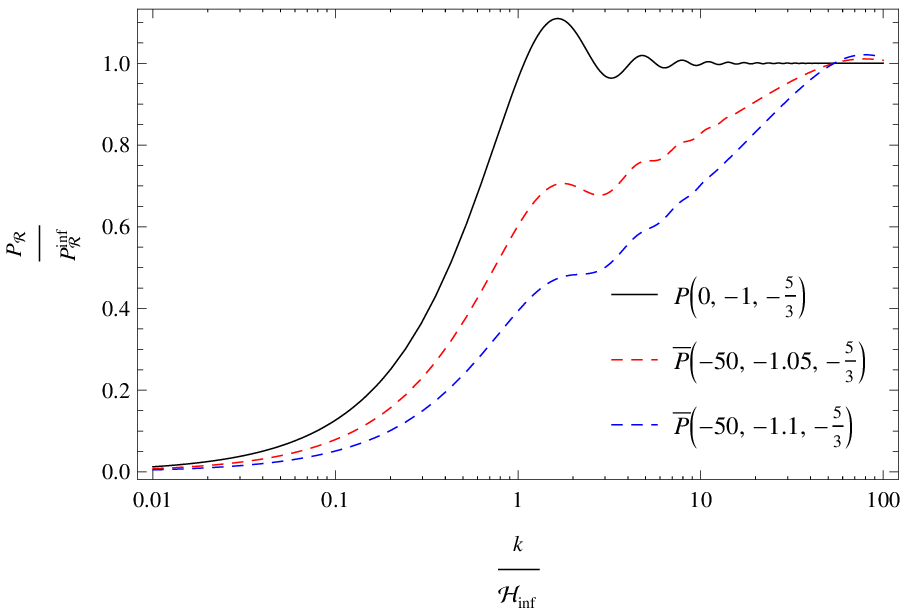}
    \end{minipage}
    \hspace{0.05cm}
\caption{$\bar{P}$ is obtained by replacing $k$ with $k/|\eta_1|$.}
    \label{fig:eei1}
\end{figure}

We plot $P(\eta_1,\omega_1,-{5}/{3})$ in Fig.\ref{fig:eei} with
different $\eta_1$ and $\omega_1$. The black solid curve, i.e.,
$P(0,-1,-{5}/{3})$, is the case with only one single
superinflationary phase before slow-roll inflation, which has been
studied in Ref.\cite{Liu:2013iha}. The perturbation mode with
wavelength $1/k=1/{\cal H}_{inf}$ is that exiting the horizon at
conformal time $\eta=0$, i.e., the onset of inflation. In
Fig.\ref{fig:eei1}, we plot $\bar{P}(\eta_1,\omega_1,-{5}/{3})$ by
replacing $k$ with ${k}/{|\eta_1|}$, which corresponds to move the
suppression of the spectrum to a smaller scale.

In Fig.\ref{fig:eei}, the effects of an intermediate phase between
superinflation and slow-roll inflation is obvious, compared with
the case without the intermediate phase. The first peaks in the
dashed curves will left shift with the increase of the duration
$|\eta_1|$ that the intermediate phase lasts.
The height of the first peak will lower with the decrease of
$\omega_1$, since the first peak corresponds to $\frac{k}{{\cal
H}_{inf}}<1$ while the spectrum is blue tilted. The power spectrum
of perturbation from the intermediate phase , i.e. phase $1$, is
roughly ${\cal P}_{{\cal R}}\sim (\frac{k}{{\cal
H}_{inf}})^{3-2v_1}$, which will be proved in Appendix B.



\subsection{The bounce inflation }\label{cei}

The pre-inflationary universe may be contracting. After the
bounce, the slow-roll inflation begins, which is the so-called
bounce inflation scenario
\cite{Piao:2003zm},\cite{Falciano:2008gt},\cite{Mielczarek:2008pf},\cite{Liu:2013kea},\cite{Biswas:2013dry},
\cite{Qiu:2014nla} with $G$-bounce, and \cite{Liu:2010fm} with
the quintom. It has been found that in this scenario the power
spectrum of primordial perturbations will get a large-scale cutoff,
which may lead to the power deficit of CMB TT-mode on a large
angular scale.

However, around the bounce the evolution of the universe is
complicated. It might be possible that after the bounce the universe
enters into an intermediate phase prior to the slow-roll
inflation. We will check how the shape of the spectrum changes with
this intermediate phase. We fixed $\omega_2=1$ and assume that the
intermediate phase is characterized by the constant state
parameter $\omega_1$, which is model dependent but satisfies
$\omega_1<-{1}/{3}$. By requiring the continuity of $|{\cal H}|$,
we get \be {\cal H}_1(0)={\cal H}_{inf} \,,\qquad {\cal
H}_2(\eta_1)=-\frac{2{ \cal H}_{inf} }{2+(1+3\omega_1)\eta_1{\cal
H}_{inf}}, \ee where ${\cal H}_{inf}$ is determined by the
amplitude of CMB fluctuations. Thus in equation set (\ref{set})
only parameters $\eta_1$ and $\omega_1$ are left to be free.

\begin{figure}[t!]
    \centering
\begin{minipage}[b]{0.45\linewidth}
    \centering
    \includegraphics[width=0.95\textwidth]{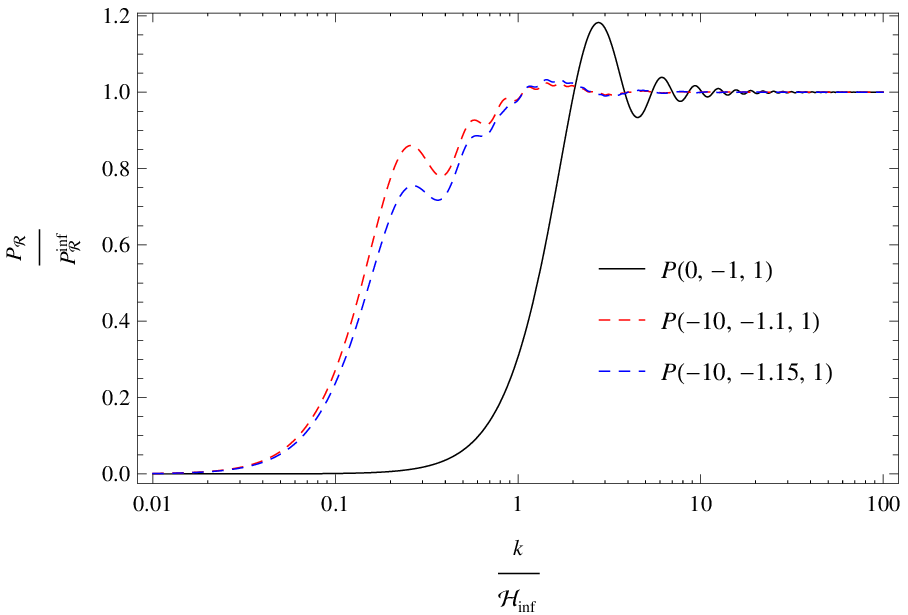}
    \end{minipage}
    \hspace{0.05cm}
\begin{minipage}[b]{0.45\linewidth}
    \centering
    \includegraphics[width=0.95\textwidth]{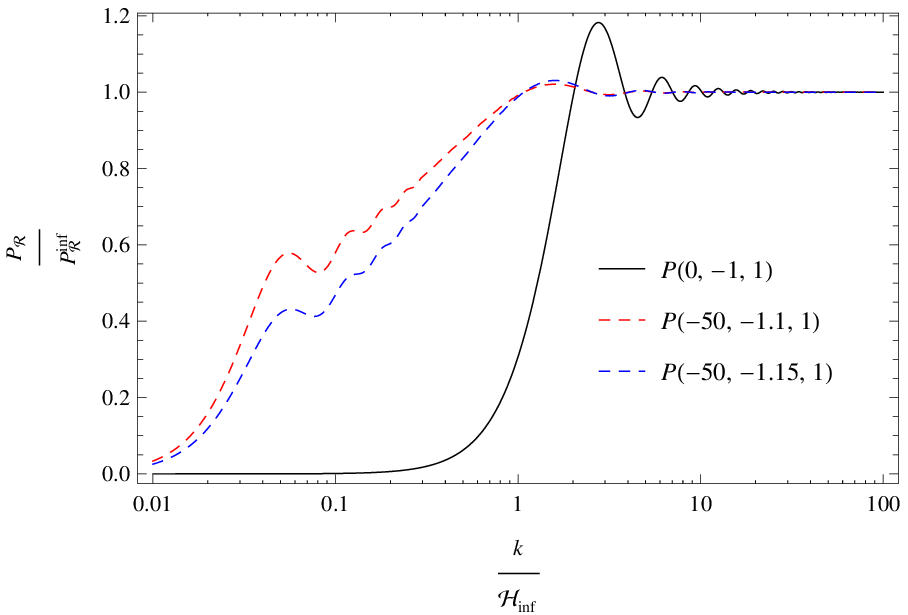}
    \end{minipage}
    \hspace{0.05cm}
\caption{ The power spectrum $P(\eta_1,\omega_1,1)$ of the model, in
which there is an intermediate phase in the bounce inflation scenario,
with different $\eta_1$ and $\omega_1$. $P(0,-1,1)$ corresponds to
the case in Ref.\cite{Liu:2013iha} where there is only one single
contracting phase before inflation. }
    \label{fig:cei}
\end{figure}
\begin{figure}[t!]
    \centering
\begin{minipage}[b]{0.45\linewidth}
    \centering
    \includegraphics[width=0.95\textwidth]{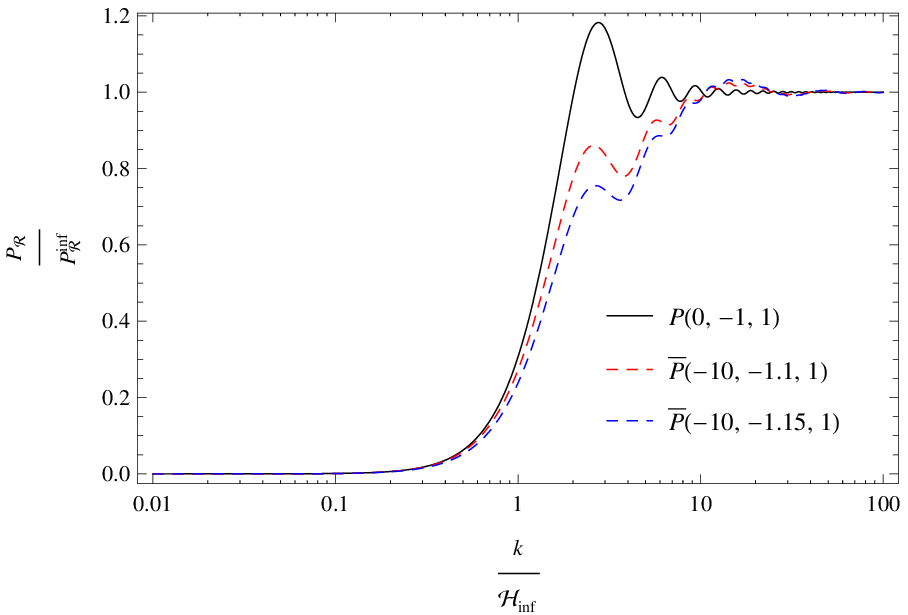}
    \end{minipage}
    \hspace{0.05cm}
\begin{minipage}[b]{0.45\linewidth}
    \centering
    \includegraphics[width=0.95\textwidth]{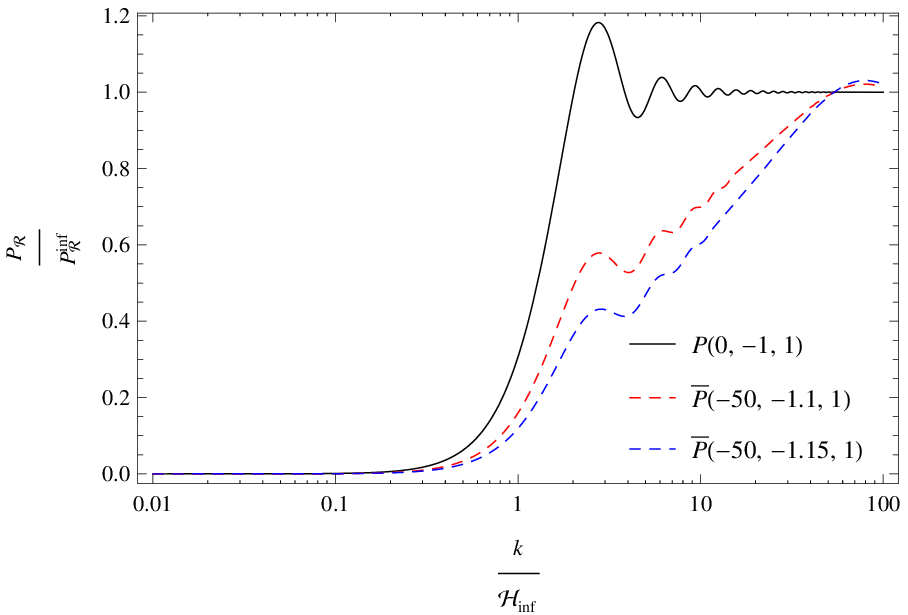}
    \end{minipage}
    \hspace{0.05cm}
\caption{$\bar{P}$ is obtained by replacing $k$ with
$k/|\eta_1|$.}
    \label{fig:cei1}
\end{figure}

We plot $P(\eta_1,\omega_1,1)$ with different $\eta_1$ and
$\omega_1$ in Fig.\ref{fig:cei}, in which the black solid curves,
i.e. $P(0,-1,1)$, correspond to the bounce inflation studied
explicitly in \cite{Piao:2003zm} \cite{Liu:2013kea}. In
Fig.\ref{fig:cei}, the effects of an intermediate phase between
contraction and slow-roll inflation is obvious. The case is
similar to \ref{eei}. In Fig.\ref{fig:cei1}, we plot
$\bar{P}(\eta_1,\omega_1,1)$ by replacing $k$ with
${k}/{|\eta_1|}$, which corresponds to move the suppression of the
spectrum to a smaller scale. Actually, similar shapes of the power
spectrum have also been obtained in Ref.\cite{Kitazawa:2014dya}.


\section{Discussion}\label{conclusion}

The large-scale power deficit in the CMB TT-mode spectrum may
imply certain pre-inflationary physics, e.g., a contracting phase
followed by the bounce or a superinflationary phase before
slow-roll inflation, which can provide a singular-free realization
of inflation. However, the physics of the pre-inflationary era
might be complex; sometimes a single phase can hardly reflect the
drastic change of the background dynamics. Thus it is interesting
to have a quantitative estimate for the power spectrum of
primordial perturbations from an arbitrary pre-inflationary era,
involving multiple phases with different background dynamics.

We perform a model-independent calculation for the power spectrum
of primordial perturbations produced during the pre-inflation with
different background evolutions.
We require the relevant physical quantities to continuously pass
through the matching surface between adjacent phases, and we obtain a
universal formula (\ref{prc1c2}) for the primordial spectrum in
terms of the recursive Bogoliubov coefficients.

We apply our formula to the bounce inflation and the
superinflation preceding slow-roll inflation  and show how the
intensity of the CMB power suppression on the large scale is affected
by the pre-inflationary physics. It is found that due to the
existence of the intermediate phase, the intensity of the power
suppression becomes model dependent.

\begin{figure}[t]
\includegraphics[width=10.0cm]{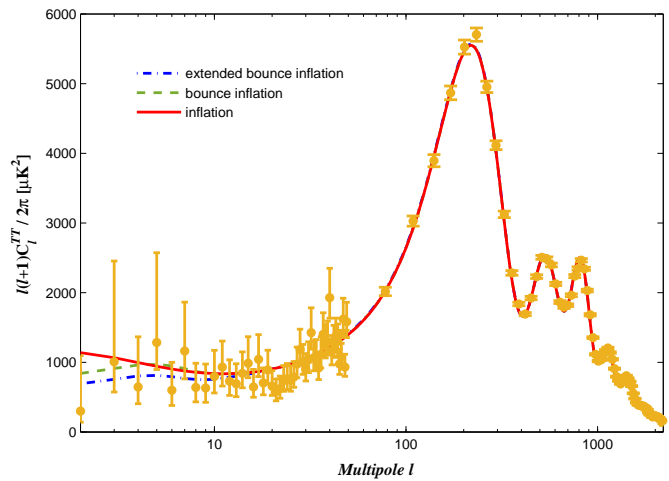}
\caption{The CMB TT-mode power spectra for inflation
with power-law power spectrum(red solid line), bounce inflation
model(green dashed line), and extended bounce inflation(blue dashed line), in
which $\Omega_bh^2=0.02203$, $\Omega_ch^2=0.1204$, $\tau=0.09$,
$100\Theta_s=1.04$, $A_s=2.1\times10^{-9}$, $n_s=0.961$. The
orange points are the Planck 2013 data with 1$\sigma$ errors. }
\label{fig:CMB}
\end{figure}

We, with the power spectrum in Fig.\ref{fig:cei}, plot the CMB TT-mode
spectrum in Fig.\ref{fig:CMB}, in which Planck 2013 data are used
and the model with an intermediate phase is called the extended
bounce inflation. It has been found in Ref.\cite{Liu:2013kea} that
the bounce inflation model can improve the fit to the data with
$\Delta\chi_{\rm eff}^2\approx -4.6$ with respect to the standard
inflation model with the power-law spectrum. Thus, it is interesting to
have a global fitting analysis with the Planck new data, which is
left in the upcoming work.

In addition, both the bounce inflation and the superinflation
preceding slow-roll inflation may also explain a large dipole
power asymmetry at low-$l$ in CMB TT-mode spectrum
\cite{Liu:2013kea},\cite{Liu:2013iha}; see also other attempts
\cite{Lyth:2013vha},\cite{McDonald:2013aca},\cite{Liddle:2013czu},\cite{Namjoo:2013fka},\cite{Mazumdar:2013yta},\cite{Cai:2013gma},
as well as \cite{Kohri:2013kqa},\cite{Kanno:2013ohv}. Actually,
there may also be dipole power asymmetry in CMB polarization
\cite{Namjoo:2014pqa},\cite{Zarei:2014raa}, which might be larger
than those in the TT-mode power spectrum. Thus it is interesting to
have an estimate for the dipole power asymmetry of CMB polarization in
the scenarios discussed here.

For the sake of simplifying the analysis and algorithm, we have
made several assumptions. We have divided the pre-inflationary
evolution into multiple phases and assumed that each phase
possesses a constant equation of state parameter, and the
transitions from one phase to the next are instantaneous. But in
reality, the equation of state parameter may be nonconstant, the
transitions should be smooth, and the equation of state parameter
should change smoothly. However, as long as there are long phases with
approximately constant equation of state parameters and relatively
quick transitions between the different phases, the
approximation we adopted to simplify our analysis and
algorithm is reasonable. Of course, the influence on the
perturbations induced by a nonconstant equation of state
parameter is interesting, and the effects of the modes that exit
the Hubble radius near the transition from one phase to another
are even more intriguing, which have been studied in
\cite{Biswas:2013lna},\cite{Biswas:2014kva}.



Throughout, we have used perturbation and background equations
that are valid for general relativity (GR), since we lack the knowledge of the new
physics that is needed for the pre-inflationary evolution.
To implement bounce and superinflation evolutions in GR, we must violate
the null energy condition(NEC) or require a closed universe. Otherwise,
one must go beyond GR.

The violation of NEC in GR usually leads to ghosts, which indicate dangerous instability. One of the several ways to avoid ghosts is by introducing the higher-order derivative scalar field. In some ghost-free Galileon bounce or superinflation models, by delicately designing the Lagrangian, one can get the perturbation equation similar to ours in the form (see, e.g., \cite{Qiu:2011cy},\cite{Liu:2011ns},\cite{Qiu:2015nha}), in which the sound speed is constant, which is set as 1 here. In that case, our analysis is still able to approximately capture the bounce or superinflation dynamics. Additionally, in singular bounce models, e.g., the original ekpyrotic model, the background is GR and the perturbation equation does not need to be modified, until the scale factor $a(t)$ is so small that quantum gravity effects become important. Then, if the bounce period is short so that the link between the contracting phase and the expanding phase is approximately instantaneous just as we have assumed, our analysis is also approximately valid. However, in some kinds of modified gravity theories, in which the perturbation equations are modified, especially when higher-order terms of gravity appear in the perturbation equations, e.g., \cite{Biswas:2014tua}, our analysis will no longer be robust and will need to be revisited.

It is significant that recently in \cite{Biswas:2015kha}, some general results for the evolutions of perturbation modes during bounce have become available, which might be used to track the perturbations during the bounce. And obtaining the recursive matrix for some specific bouncing model to better understand how the modes evolve during this period is an interesting issue, which we will back to in the future.


We have assumed $c_s^2=1$ after Eq.(\ref{u}), which is
suitable for the cosmology driven by the scalar field. However, for
an ideal fluid, $c_s^2=\omega$, and for $\omega < 0$, it is
negative which indicates an instability in the system. It is well
known in thermal physics that a negative heat capacity typically
implies that one is looking at thermal fluctuations around an
incorrect (possibly tachyonic) vacuum. For $\omega < 0$,
essentially the short wavelength fluctuations grow exponentially
rather than oscillate.  Actually, for scalar fields, $c_s^2$
may also be different from $1$. The effects of varying sound speed
on primordial perturbations has been studied; see,
e.g., \cite{Nakashima:2010sa},\cite{Firouzjahi:2014fda}. But as long
as $c_s^2$ is a positive constant during each phase, our result is
not qualitatively altered.

When deriving (\ref{prc1c2}), we assume that around and at
matching surfaces the perturbation mode has no ghost
and gradient instabilities. How to implement such a requirement
is an interesting issue,
e.g., \cite{Pirtskhalava:2014esa},\cite{Battarra:2014tga}. However,
if this requirement is not satisfied, which actually is not
allowed physically, the result of the perturbation spectrum will
be strongly affected by the physical details around the matching
surface.

\textbf{Acknowledgments}

We thank Zhi-Guo Liu for helpful discussions. This work is
supported by NSFC, No.11222546, and National Basic Research
Program of China, No.2010CB832804. We acknowledge the use of CAMB.

\section*{Appendix A: The Wronskian constraint for $C_{i,1}$ and $C_{i,2}$ }

Initially $k\ll {\cal H}$, and the perturbation should be in its
minimal energy state, i.e., BD vacuum. Thus, the BD vacuum is
usually regarded as the initial state of the primordial
perturbations. The effects of non-BD initial states on the
primordial perturbation has been discussed in, e.g.,
Refs., \cite{Holman:2007na},\cite{Dey:2011mj},\cite{Gong:2013yvl},\cite{Agarwal:2012mq},\cite{Ashoorioon:2013eia},\cite{Chen:2014vja},\cite{Kundu:2011sg},\cite{Kundu:2013gha}.
Here, the Bogoliubov coefficients are \ba
\alpha_i=2\sqrt{\frac{k}{\pi}} C_{i,1}, \nonumber\\
\label{bogoliubov} \beta_i=2\sqrt{\frac{k}{\pi}} C_{i,2}. \ea The
Wronskian (or canonical normalization) constraint requires
$|\alpha_i|^2-|\beta_i|^2=1$. The initial state, i.e., BD vacuum,
corresponds to $\alpha_{i_{max}}=1$ and $\beta_{i_{max}}=0$, i.e.,
\be C_{i_{max},1}=\frac{\sqrt{\pi}}{2\sqrt{k}}, \qquad
C_{i_{max},2}=0. \ee

We will prove that if $C_{i_{max},1}$ and $C_{i_{max},2}$ satisfy
the Wronskian constraint, for any phase $i$, $C_{i,1}$ and
$C_{i,2}$ also satisfy this constraint. This is equivalent to the
statement that for any phase $i$, if \be \frac{4k}{\pi}\left(
|C_{i+1,1}|^2-|C_{i+1,2}|^2 \right)=1 \label{proof1}\ee is
satisfied, we always have \be \frac{4k}{\pi}\left(
|C_{i,1}|^2-|C_{i,2}|^2 \right)=1. \label{proof}\ee

It equals the proof that \ba \left(\begin{array}{ccc} \overline{
C}_{i,1} & \overline{C}_{i,2}
\end{array} \right)\times \left(
\begin{array}{ccc}1&0\\0&-1 \end{array}
\right)\times\left(\begin{array}{ccc} C_{i,1}\\C_{i,2} \end{array}
\right)= \left(\begin{array}{ccc} \overline{ C}_{i+1,1} &
\overline{C}_{i+1,2}   \end{array} \right)\times \left(
\begin{array}{ccc}1&0\\0&-1 \end{array}
\right)\times\left(\begin{array}{ccc} C_{i+1,1}\\C_{i+1,2}
\end{array} \right), \label{cm0} \ea where $\overline{C}$ is
the complex conjugation of $C$. Because of \ba
\left(\begin{array}{ccc} C_{i,1}\\C_{i,2} \end{array}\right)={\cal
M}^{(i,i+1)}\times\left(\begin{array}{ccc} C_{i+1,1}\\C_{i+1,2}
\end{array}\right), \label{cm1} \ea we have \ba  \left(
\begin{array}{ccc} \overline{ C}_{i,1} &  \overline{C}_{i,2}
\end{array} \right) &=& \left(\begin{array}{ccc} C_{i,1}\\C_{i,2}
\end{array}\right)^{\dag} = \left(\begin{array}{ccc}
C_{i+1,1}\\C_{i+1,2} \end{array}\right)^{\dag}\times \left({\cal
M}^{(i,i+1)}\right)^{\dag}
\label{cm2}\nonumber\\
&=& \left(\begin{array}{ccc} \overline{C}_{i+1,1} &
\overline{C}_{i+1,2} \end{array}\right) \times \left({\cal
M}^{(i,i+1)}\right)^{\dag}. \ea Thus after substituting
Eqs.(\ref{cm1}) and (\ref{cm2}) into Eq.(\ref{cm0}), we have \ba
\left({\cal M}^{(i,i+1)}\right)^{\dag}\times \left(
\begin{array}{ccc} 1& 0
\\0&-1  \end{array}\right) \times \left({\cal M}^{(i,i+1)}\right)
=\left( \begin{array}{ccc} 1& 0 \\0&-1  \end{array}\right), \ea
where ${\cal M}^{(i,i+1)}$ is defined in Eq.(\ref{ditui}).

Therefore, it is left to prove \ba
 \left\{ \begin{array}{ll}
|{\cal M}_{11}|^2-|{\cal M}_{21}|^2=1 \\ \label{pfc}
\overline{ {\cal M}}_{11}\times {\cal M}_{12}-\overline{ {\cal M} }_{21}\times {\cal M}_{22}=0\\
\overline{ {\cal M}}_{12}\times {\cal M}_{11}-\overline{ {\cal M} }_{22}\times {\cal M}_{21}=0\\
|{\cal M}_{12}|^2-|{\cal M}_{22}|^2=-1
\end{array}\right. .
\ea This actually can be obtained by using \be \overline{H^{(1)}_v
}(\xi)=H^{(2)}_v(\xi), \ee and \be
H^{(1)}_v(\xi)H^{(2)}_{-1+v}(\xi)-H^{(1)}_{-1+v}(\xi)H^{(2)}_v(\xi)=-\frac{4i}{\pi
\xi}.\ee  Hence, for any phase $i$, if (\ref{proof1}) is
satisfied, we always have (\ref{proof}).

The Wronskian constraint for $C_{0,1}$ and $C_{0,2}$ is \be
\frac{4k}{\pi}\left( |C_{0,1}|^2-|C_{0,2}|^2 \right)=1. \ee Thus,
${\cal P}_{\cal R}$ in (\ref{prc1c2}) becomes \be {\cal P}_{\cal
R}={\cal P}^{inf}_{\cal
R}\frac{4}{\pi}k\left|C_{0,1}-C_{0,2}\right|^2={\cal
P}^{inf}_{\cal R}\left|\alpha_{0}-\beta_{0}\right|^2, \ee where
$\beta_{0}\neq 0$. Thus, in a certain sense the scenario with the
pre-inflationary era is equivalent to the inflation scenario with
the non-BD initial state.

\section*{Appendix B: The perturbation spectrum from pre-inflationary phase $i$}\label{appendix}

In this appendix, we will prove that the power spectrum of
perturbations produced in phase $i$ is approximately \be {\cal
P}_{{\cal R}}\sim \left( \frac{k}{ {\cal H}_{inf} }
\right)^{n_i-1}, \label{app1}\ee where \be n_i-1=3-2v_i,
\label{app2}\ee and $v_i$ is defined as $
v_i=\frac{3}{2}\left|\frac{1-\omega_i}{1+3\omega_i}\right|$.

First, we give a proof for the case with $i_{max}=1$. The wave
number of the perturbations produced in phase $i$ satisfies $k\ll
{\cal H}_{inf}$. We assume that phase $1$ is an expanding phase,
thus $\omega_1<-{1}/{3}$. According to Sec.\ref{au}, we have \ba
u_0&=&\sqrt{-k\eta+\frac{k}{ {\cal H}_{inf} } }\left( C_{0,1}H^{(1)}_{3/2}(-k\eta+\frac{k}{ {\cal H}_{inf} })+C_{0,2}H^{(2)}_{3/2}(-k\eta+\frac{k}{ {\cal H}_{inf} })   \right),\nonumber\\
u_1&=&\sqrt{-k\eta-\frac{2k}{ (1+3\omega_1){\cal H}_{inf} } }\Big( C_{1,1}H^{(1)}_{v_i}(-k\eta-\frac{2k}{ (1+3\omega_1){\cal H}_{inf} })\nonumber\\
&\,\,&\qquad\qquad\qquad\qquad\qquad+C_{1,2}H^{(2)}_{v_i}(-k\eta-\frac{2k}{
(1+3\omega_1){\cal H}_{inf} })   \Big), \ea where \be
C_{1,1}=\frac{\sqrt{\pi}}{2 \sqrt{k} }, \qquad\qquad C_{1,2}=0.
\ee Because ${\cal H}_{inf}$ always appears with $k$ as
${k}/{{\cal H}_{inf}}$ below, we will set ${\cal H}_{inf}=1$ for
convenience, and actually can easily get it back. We have \ba
{\cal P}_{\cal R}={\cal P}^{inf}_{\cal
R}\frac{4}{\pi}k\left|C_{0,1}-C_{0,2} \right|^2, \ea where

\ba
C_{0,1}&=&-\frac{\pi k}{8i}\Bigg\{ \sqrt{ \frac{-\pi}{(1+3\omega_1)(2k)^3}  }\bigg[ 2k\Big( -H^{(1)}_{-1+v_1}(\frac{-2k}{1+3\omega_1})+H^{(1)}_{1+v_1}(\frac{-2k}{1+3\omega_1})  \Big)H^{(2)}_{3/2}(k)\nonumber\\
&\,\,&+H^{(1)}_{v_1}(\frac{-2k}{1+3\omega_1}) \Big( 2kH^{(2)}_{1/2}(k)+3(1+\omega_1)H^{(2)}_{3/2}(k)-2kH^{(2)}_{5/2}(k)  \Big)
\bigg]
\Bigg\},
\nonumber\\
\ea
\ba
C_{0,2}&=&\frac{\pi k}{8i}\Bigg\{ \sqrt{ \frac{-\pi}{(1+3\omega_1)(2k)^3} }\bigg[ 2k\Big( H^{(1)}_{1/2}(k)-
H^{(1)}_{5/2}(k) \Big)H^{(1)}_{v_1}(\frac{-2k}{1+3\omega_1})
\\\nonumber
&\,\,&+H^{(1)}_{3/2}(k)\Big( -2kH^{(1)}_{-1+v_1}(\frac{-2k}{1+3\omega_1})+3(1+\omega_1)H^{(1)}_{v_1}(\frac{-2k}{1+3\omega_1})+
2k H^{(1)}_{1+v1}(\frac{-2k}{1+3\omega_1})  \Big)
\bigg]
\Bigg\}.
\ea

The special case is $\omega_1=-{5}/{3}$, which gives $v_1=1$,
$-1+v_1=0$. When $k\ll1$, i.e. $\frac{k}{{\cal H}_{inf}}\ll 1$, we get
\be {\cal P}_{\cal R}\approx{\cal P}_{\cal R}^{inf}\frac{4k}{\pi
{\cal H}_{inf} }\left( 1+\frac{k^2}{12{\cal H}_{inf}^2}\ln{\frac{k}{4{\cal
H}_{inf} }}-\frac{k^2}{6{\cal H}_{inf}^2} \right)^2\sim \frac{k}{{\cal
H}_{inf}  }. \ee

If $-{5}/{3}<\omega_1<-{1}/{3}$, then $v_1>1$, $-1+v_1>0$. When
$\frac{k}{{\cal H}_{inf}}\ll 1$, we get \ba {\cal P}_{\cal R}
&\approx& {\cal P}_{\cal
R}^{inf}\frac{1}{\pi}\left(\frac{-1}{1+3\omega_1}
\right)^{3-2v_1}\cdot \left(\frac{k}{{\cal H}_{inf}} \right)^{3-2v_1}
\nonumber\\
&\,\,&\times\Big\{-(1+3\omega_1)\Gamma(v_1)+\frac{k^2}{6{\cal H}_{inf}}\Big[(1+3\omega_1)\Gamma(v_1)-\Gamma(-1+v_1)
\Big]
\Big\}^2
\nonumber\\
&\sim& \left(\frac{k}{{\cal H}_{inf}} \right)^{3-2v_1}.
\ea

If $\omega_1<-{5}/{3}$, then $v_1<1$, $-1+v_1<0$,
$H^{(1)}_{-1+v_1}$ should be replaced by
$e^{i(-1+v_1)\pi}H^{(1)}_{1-v_1}$. When $\frac{k}{{\cal H}_{inf}}\ll
1$, we get \ba {\cal P}_{\cal R} &\approx& {\cal P}_{\cal
R}^{inf}\frac{1}{\pi}\left(\frac{k}{{\cal H}_{inf}} \right)^{3-2v_1}
\nonumber\\
&\,\,&\times\left[\Gamma^2(v_1)(-1-3\omega_1)^{-1+2v_1}+\frac{1}{3}\left(\frac{k}{{\cal H}_{inf}} \right)^{2v_1}\cos(v_1\pi)\Gamma(1-v_1)\Gamma(v_1)\right]
\nonumber\\
&\sim& \left(\frac{k}{{\cal H}_{inf}} \right)^{3-2v_1}.
\ea

Thus, we always have ${\cal P}_{{\cal R}}\sim\left( \frac{k}{ {\cal
H}_{inf} } \right)^{n_i-1}$. This result can also apply to the case
with $\omega_1>-{1}/{3}$.

Then, we focus on the case with $i_{max}>1$. Though the spectra
index seems nontrivial, based on some assumptions, we could get a
similar result.

We assume that the pre-inflationary era is phase $\tilde{i}$
dominated, i.e., the phase $\tilde{i}$ lasts long enough, nearly,
$(\eta_{\tilde{i}-1}-\eta_{\tilde{i}})\rightarrow\infty$. The
modes of the perturbations produced in phase $\tilde{i}$ satisfy
${\cal H}_{\tilde{i} }(\eta_{\tilde{i} })\ll k \ll {\cal
H}_{\tilde{i} }(\eta_{\tilde{i}-1})$. Thus, for $i\geq \tilde{i}$,
we always have $x\gg 1$ and $y\gg 1$, $x$ and $y$ are those in
Eq.(\ref{rec}), and we obtain \ba {\cal M}_{11}&=&\left(
1-\frac{x^{-1}-y^{-1}}{4i} \right) e^{-i(x-y-\frac{ v_i-v_{i+1}
}{2}\pi)}\approx e^{-i(x-y-\frac{ v_i-v_{i+1} }{2}\pi)},
\label{mm}
\nonumber\\
{\cal M}_{12}&=&-\frac{x^{-1}-y^{-1}}{4}e^{ -i(x+y-\frac{v_i+v_{i+1}}{2}\pi )   }\approx0,
\nonumber\\
{\cal M}_{21}&=&-\frac{x^{-1}-y^{-1}}{4}e^{ i(x+y-\frac{v_i+v_{i+1}}{2}\pi )   }\approx0,
\nonumber\\
{\cal M}_{22}&=&\left( 1+\frac{x^{-1}-y^{-1}}{4i} \right) e^{i(x-y-\frac{ v_i-v_{i+1} }{2}\pi)}\approx e^{i(x-y-\frac{ v_i-v_{i+1} }{2}\pi)}.
\ea
Thus, when $i\geq \tilde{i}$, the recursive matrixes ${\cal M}^{(i,i+1)}$ are diagonal. We get
\ba
\left( \begin{array}{ccc} C_{\tilde{i},1}\\C_{\tilde{i},2}  \end{array}\right)
&=&{\cal M}^{(\tilde{i},\tilde{i}+1)}\times{\cal M}^{(\tilde{i}+1,\tilde{i}+2)}\times\cdots\times{\cal M}^{(i_{max}-1,i_{max})}\times
\left(\begin{array}{ccc} C_{i_{max},1} \\ C_{i_{max},2}  \end{array}\right)
\nonumber\\
&=&\left(\begin{array}{ccc} e^{-i\cdot E} & 0 \\ 0 & e^{i\cdot E}
\end{array}\right) \times\left(\begin{array}{ccc}\frac{\sqrt{\pi}
}{2\sqrt{k}} \\0 \end{array}\right) =e^{-i\cdot
E}\left(\begin{array}{ccc} C_{i_{max},1} \\ 0  \end{array}\right),
\ea where $E$ is real,
$C_{i_{max},1}=\frac{\sqrt{\pi}}{2\sqrt{k}}$. Because  we are
concerned about $\left| C_{0,1}-C_{0,2} \right|$ in the final result,
$e^{-i \cdot E}$ makes no difference, we can ignore it and write
that \ba \left( \begin{array}{ccc}
C_{\tilde{i},1}\\C_{\tilde{i},2}
\end{array}\right)=\frac{\sqrt{\pi}}{2\sqrt{k}} \left(
\begin{array}{ccc}1\\0  \end{array} \right)=\left(
\begin{array}{ccc}C_{i_{max},1}\\0  \end{array} \right).
\label{ctil} \ea


For $i<\tilde{i}$, we always have $x\ll 1$, $y\ll1$. By expanding
the Hankel functions to second order, we have \ba
C_{i,1}+C_{i,2}&=&({\cal M}_{11}+{\cal M}_{21} )C_{i+1,1}+({\cal
M}_{12}+{\cal M}_{22})C_{i+1,2} \label{ci1ci2}
\nonumber\\
&\approx& A_{i,1}\cdot(C_{i+1,1}+C_{i+1,2})\left(\frac{k}{ {\cal H}_i(\eta_i)  } \right)^{v_{i+1}-v_i  }
\nonumber\\
&+&A_{i,2}\cdot (C_{i+1,1}-C_{i+1,2})\left(\frac{k}{ {\cal H}_i(\eta_i)}    \right)^{-v_{i+1}-v_i},
\nonumber\\
C_{i,1}-C_{i,2}&=&({\cal M}_{11}-{\cal M}_{21} )C_{i+1,1}+({\cal M}_{12}-{\cal M}_{22})C_{i+1,2}
\nonumber\\
&\approx& A_{i,3}\cdot(C_{i+1,1}+C_{i+1,2})\left(\frac{k}{ {\cal H}_i(\eta_i)  } \right)^{v_{i+1}+v_i  }
\nonumber\\
&+&A_{i,4}\cdot (C_{i+1,1}-C_{i+1,2})\left(\frac{k}{ {\cal
H}_i(\eta_i)}    \right)^{-v_{i+1}+v_i}, \ea where the constants
$A_{i,1}, A_{i,2},A_{i,3},A_{i,4}$ are independent with $k$, and
$v_i$ is defined as $
v_i=\frac{3}{2}\left|\frac{1-\omega_i}{1+3\omega_i}\right|$. Because $k\ll {\cal H}( \eta_{
\tilde{i}-1} )<{\cal H}(\eta_{ \tilde{i}-2 })<\cdots<{\cal
H}(\eta_0)$, we will not care about ${\cal H}( \eta_{\tilde{i}-1} ),
..., {\cal H}(\eta_0)$ below, and just regard $\frac{k}{ {\cal
H }(\eta_i) } \ll 1$ as $k\ll 1$ instead.

According to Eqs.(\ref{ctil}) and  (\ref{ci1ci2}), we get
\ba
C_{ \tilde{i}-1,1}+C_{ \tilde{i}-1,2 }&=&C_{i_{max},1 }\cdot\left( A_{\tilde{i}-1,1}\cdot k^{ v_{\tilde{i}} -v_{\tilde{i}-1} }+A_{ \tilde{i}-1,2 }\cdot k^{ -v_{\tilde{i}}-v_{\tilde{i}-1 } }  \right),
\nonumber\\
C_{ \tilde{i}-1,1}-C_{ \tilde{i}-1,2 }&=&C_{i_{max},1 }\cdot\left( A_{\tilde{i}-1,3}\cdot k^{ v_{\tilde{i}} +v_{\tilde{i}-1} }+A_{ \tilde{i}-1,4 }\cdot k^{ -v_{\tilde{i}}+v_{\tilde{i}-1 } }  \right),
\ea
and then
\ba
C_{ \tilde{i}-2,1}+C_{ \tilde{i}-2,2 }&=&C_{i_{max},1 }\cdot\left( A_{\tilde{i}-2,1}\cdot k^{ v_{\tilde{i}} -v_{\tilde{i}-2} }+A_{ \tilde{i}-2,2 }\cdot k^{ -v_{\tilde{i}}-v_{\tilde{i}-2 } }  \right),
\nonumber\\
C_{ \tilde{i}-2,1}-C_{ \tilde{i}-2,2 }&=&C_{i_{max},1 }\cdot\left( A_{\tilde{i}-2,3}\cdot k^{ v_{\tilde{i}} +v_{\tilde{i}-2} }+A_{ \tilde{i}-2,4 }\cdot k^{ -v_{\tilde{i}}+v_{\tilde{i}-2 } }  \right),
\nonumber
\ea
\\
$$\vdots\qquad\qquad\vdots\qquad\qquad\vdots\qquad\qquad\vdots$$
\ba
C_{0,1}+C_{0,2}&=& C_{i_{max},1}\cdot \left( A_{0,1}\cdot k^{v_{\tilde{i}}-v_0}+A_{0,2}\cdot k^{ -v_{\tilde{i}}-v_0  } \right),
\nonumber\\
C_{0,1}-C_{0,2}&=& C_{i_{max},1}\cdot \left( A_{0,3}\cdot k^{v_{\tilde{i}}+v_0}+A_{0,4}\cdot k^{ -v_{\tilde{i}}+v_0  } \right).
\ea

After neglecting the high order terms, we have \ba
C_{0,1}-C_{0,2}\approx C_{ i_{max},1 }\cdot A_{0,4}\cdot
k^{-v_{\tilde{i}}+v_0  }\sim \frac{ \sqrt{\pi} }{ 2\sqrt{k}
}k^{-v_{\tilde{i}}+v_0 }, \ea where $v_0={3}/{2}$ for the
inflationary phase. Therefore \ba {\cal P}_{\cal R}={\cal
P}^{inf}_{\cal R}\frac{4}{\pi}k\left|C_{0,1}-C_{0,2} \right|^2
\sim{\cal P}^{inf}_{\cal R}\cdot k^{3-2v_{\tilde{i}} }.
\label{Pappro}\ea It should be noted that (\ref{Pappro}) is a good
approximation only if phase $\tilde{i}$ lasts long enough,
i.e., ${\cal H}_{ \tilde{i} }(\eta_{\tilde{i}})\ll k\ll {\cal
H}_{\tilde{i}}(\eta_{ \tilde{i}-1 })$. The reason is apparent,
since if   phase $\tilde{i}$ lasts only a very short time, the
oscillations of spectrum around the matching surface would destroy
the relation we proofed above.

\begin{figure}[htbp]
\includegraphics[scale=2,width=0.5\textwidth]{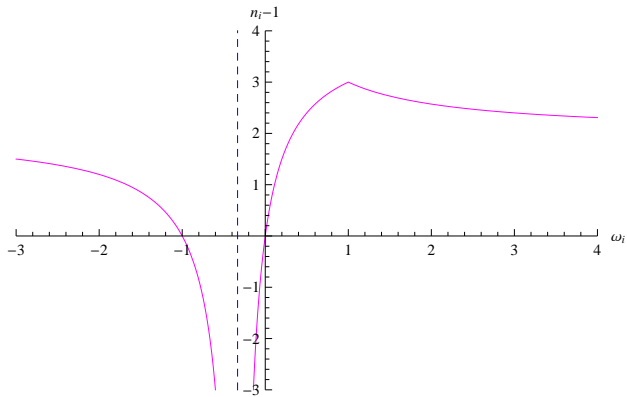}
\caption{The figure of $n_i-1$ with respect to $\omega_i$, which
is plotted based on $n_i-1=3-2v_i$ and $
v_i=\frac{3}{2}\left|\frac{1-\omega_i}{1+3\omega_i}\right|$. The
magenta curve is $n_i-1$. And the blue dashed line is
$\omega_i=-{1}/{3}$.} \label{fig:v1}
\end{figure}

We plot $n_i-1$ with respect to $\omega_i$ in Fig.\ref{fig:v1}
(see also \cite{Piao:2004jg}), which provides guidance for
building a pre-inflationary model leading to a cutoff spectrum on a
large scale. We see that the power spectrum is strong blue-tilt
only for the expansion with $\omega_i<-1$ and the contraction with
$\omega_i>0$.

\end{document}